\documentclass{osa-article}

\usepackage{amsmath}
\usepackage{graphicx}
\usepackage{subcaption}
\DeclareCaptionLabelFormat{nocaption}{}

\journal{oe}

\articletype{Research Article}

\begin{document}

\title{Single-frame far-field diffractive imaging with randomized illumination}

\author{Abraham L. Levitan,\authormark{1,*} Kahraman Keskinbora,\authormark{1,2,3} Umut T. Sanli,\authormark{2} Markus Weigand,\authormark{4} and Riccardo Comin\authormark{1}}

\address{\authormark{1} Massachusetts Institute of Technology, 77 Mass Avenue, Cambridge, MA, USA\\
\authormark{2} Max Planck Institute for Intelligent Systems, Heisenbergstrasse 3, 70569, Stuttgart, Germany\\
\authormark{3} Harvard University, John A. Paulson School of Applied Sciences, Center for Integrated Quantum Materials, 19 Oxford Street, Cambridge, MA, USA\\
\authormark{4} Helmholtz-Zentrum Berlin f{\"u}r Materialien und Energie GmbH, 12489 Berlin, Germany}

\email{\authormark{*}alevitan@mit.edu} 



\begin{abstract}
We introduce a single-frame diffractive imaging method called Randomized Probe Imaging (RPI). In RPI, a sample is illuminated by a structured probe field containing speckles smaller than the sample's typical feature size. Quantitative amplitude and phase images are then reconstructed from the resulting far-field diffraction pattern. The experimental geometry of RPI is straightforward to implement, requires no near-field optics, and is applicable to extended samples. When the resulting data are analyzed with a complimentary algorithm, reliable reconstructions which are robust to missing data are achieved. To realize these benefits, a resolution limit associated with the numerical aperture of the probe-forming optics is imposed. RPI therefore offers an attractive modality for quantitative X-ray phase imaging when temporal resolution and reliability are critical but spatial resolution in the tens of nanometers is sufficient. We discuss the method, introduce a reconstruction algorithm, and present two proof-of-concept experiments: one using visible light, and one using soft X-rays.
\end{abstract}


\section{Introduction}
Diffractive imaging refers to a collection of computational imaging techniques that reconstruct quantitative amplitude and phase images directly from diffraction patterns \cite{Chapman2010,Miao2015}. Twenty years after the first demonstrations, \cite{Miao1999,Marchesini2003,Shapiro2005,Chapman2006} this methodology has spurred a wave of technical and scientific innovations and motivated the development of a new generation of advanced X-ray light sources. Today, diffractive imaging offers unprecedented opportunities for quantitative phase imaging using both single-frame and multi-frame methods.

In general, single-frame methods can perform time-resolved imaging of non-reproducible dynamics but multi-frame methods are more flexible and reliable. The comparison between Coherent Diffractive Imaging (CDI) and ptychography exemplifies this compromise. CDI, like most single-frame methods, uses small samples with a sharply defined boundary. The boundary is typically imposed by an opaque mask deposited on the sample \cite{Miao2015}. This requirement ultimately restricts what systems can be studied and predefines the field of view. In contrast, ptychography uses multiple diffraction patterns from overlapping regions of a sample to improve the  reconstruction's reliability and allow extended samples to be imaged \cite{Rodenburg2004,Thibault2008,Guizar-Sicairos2008,Maiden2009,Thibault2009,Stockmar2013,Pfeiffer2018}. This trade-off has driven an enduring search for single-frame imaging methods which retain the reliability and flexibility of ptychography. Here we implement an approach to tackling this challenge which synthesizes two ingredients from different corners of the diffractive imaging world.

The first ingredient is Band-Limited Random (BLR) illumination, a variety of structured illumination which has been explored in the context of ptychographic imaging \cite{Maiden2013,Marchesini2016,Morrison2018,Marchesini2019,Odstrcil2019}. Our choice to use this illumination arises from the longstanding observation that the reliability of a phase retrieval problem is typically improved by the presence of high-frequency phase structures \cite{Guizar-Sicairos2012}. This idea has led to a number of proposed diffractive imaging methods which use randomized illumination \cite{Fannjiang2012no2,Fannjiang2012,Horisaki2016,Horisaki2017} or a quadratic phase ramp as in Fresnel CDI\cite{Williams2006, Abbey2008, Williams2010}. The concept of using randomized illumination is also closely related to Coherent Modulation Imaging (CMI) \cite{Zhang2016,Dong2018,Ulvestad2018,Tang2019}, a method which imprints randomized high-frequency structures on the wavefield with a mask placed downstream from the sample. In this work we point out that there is strong reason to believe that, when the object function is band-limited to a sufficiently low frequency, diffraction data resulting from many of these methods can support a unique solution to the phase retrieval problem even without a finite support constraint.

This observation leads naturally to the second ingredient, a band-limiting constraint applied to the object function during reconstruction. This constraint is loosely inspired by single-shot ptychography algorithms \cite{Pan2013,Sidorenko2016,Zhou2018,He2018}. In single shot ptychography, an entire ptychography dataset is taken in a single frame by converging a grid of probes onto the sample and imaging all the diffraction patterns simultaneously. To avoid interference between neighboring diffraction patterns, the Fourier transform of the object must fall to zero beyond some band-limiting frequency. To acknowledge this band-limiting requirement on the underlying object function, single-shot ptychography algorithms explicitly impose a band-limiting constraint on the object via a judicious choice of the pixel pitch in the object array. These algorithms can produce remarkably robust reconstructions, with reliability in the same class as ptychography. It therefore seems plausible that a similar constraint could also pay dividends when analyzing data arising from experiments, such as those discussed above, where the diffraction data alone provides a particularly strong constraint when the object is band-limited.

Randomized Probe Imaging (RPI) synthesizes these two ingredients by applying a band-limiting constraint to diffraction data collected under BLR illumination. The exact structure of the illumination is found with an initial ptychography scan used for calibration. As we show, this combination provides enough information to solve the phase retrieval problem without the need for either a finite support constraint or multiple diffraction patterns. The result is a reliable single-frame (and therefore potentially single-shot) technique which can be easily implemented at most ptychography beamlines by simply replacing an optic. As a bonus, RPI is applicable to extended samples and is robust to missing regions of data, such as those caused by dead pixels or the presence of beamstops. RPI is therefore an attractive alternative to holography, CDI, and CMI in a variety of situations, such as x-ray microscopy applications where reliability is paramount, sample environments are bulky and a resolution limit in the tens of nanometers is acceptable.

\section{Randomized probe imaging}

\subsection{Basic Principles}

\begin{figure}
\captionsetup[subfigure]{labelformat=nocaption}
\centerline{\includegraphics[width=\textwidth]{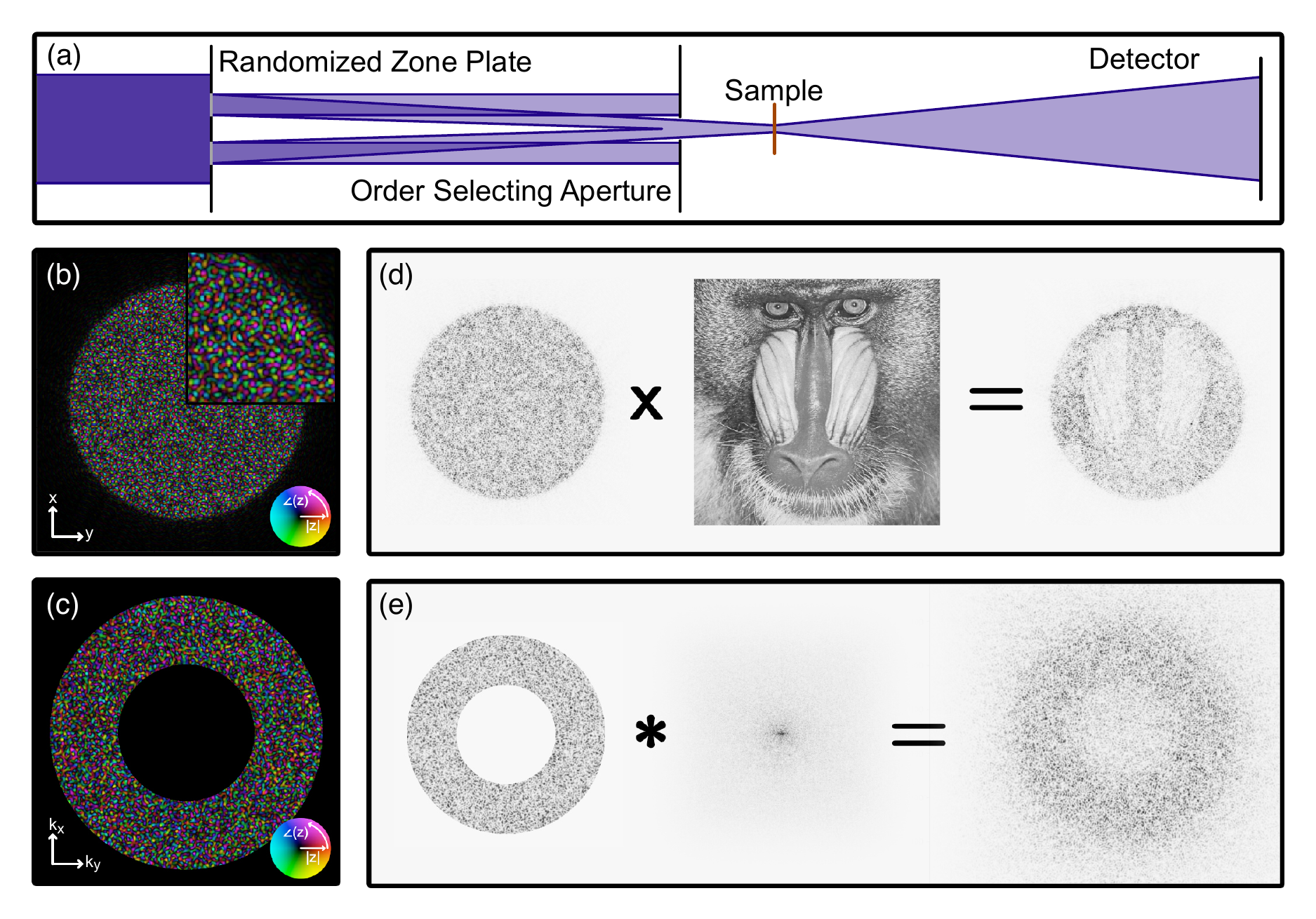}}
\begin{subfigure}{0\linewidth}
\caption{}\label{fig:exp:diagram}
\end{subfigure}
\begin{subfigure}{0\linewidth}
\caption{}\label{fig:exp:blrreal}
\end{subfigure}
\begin{subfigure}{0\linewidth}
\caption{}\label{fig:exp:blrfour}
\end{subfigure}
\begin{subfigure}{0\linewidth}
\caption{}\label{fig:exp:intreal}
\end{subfigure}
\begin{subfigure}{0\linewidth}
\caption{}\label{fig:exp:intfour}
\end{subfigure}
\caption{\textbf{Conceptual Model}. (a), A diagram outlining the experimental geometry. (b), A typical band-limited random focal spot in real space, with an inset showing the details of the finely structured amplitude and phase. (c), The corresponding representation in Fourier space. (d), The interaction model in real space, where the incident light $P(x,y)$ is multiplied by the object $O(x,y)$ to find the exit wave. (e), The same interaction model in Fourier space, where the incident illumination's Fourier transform $\tilde{P}(k_x,k_y)$ is convolved with the Fourier transform of the object $\tilde{O}(k_x,k_y)$ to find the diffraction pattern.} \label{fig:explanation}
\end{figure}

The experimental geometry at the heart of RPI, outlined in Figure \ref{fig:exp:diagram}, is the same as a geometry commonly used for ptychography. In fact, the first step of any RPI experiment is to collect a ptychography dataset to solve for the illumination wavefield. A probe-forming diffractive hologram known as a Randomized Zone Plate (RZP) focuses light of a wavelength $\lambda$ to a focal spot over a focal distance $f(\lambda)$. An Order Selecting Aperture (OSA) is used in conjunction with a central beamstop on the probe-forming optic to block all but the intended diffractive order from reaching the sample.

The only nonstandard requirement in our geometry is the use of an RZP as the probe-forming optic in place of the traditional Fresnel zone plate or pinhole. As shown in Figures \ref{fig:exp:blrreal} and \ref{fig:exp:blrfour} and further explored in Figure S2, distortions in this zone plate broaden the focal spot and fill it with speckles. The focal spot diameter can be controlled over a wide range during the optic design process, as discussed in Supplementary Section 2. The diameter of this speckle-filled focal spot ultimately determines the field of view of the RPI reconstructions. The speckle size, related to the numerical aperture of the zone plate, sets the resolution element. The effect of this randomized wavefield on the final diffraction pattern is very similar to the effect of the phase modulator used in CMI.

This BLR probe wavefield $P(x,y)$ then interacts with a thin sample described by an object function $O(x,y)$. The resulting exit wave $E(x,y)=P(x,y)O(x,y)$ propagates into the far field where its intensity is imaged onto an area detector (Fig \ref{fig:exp:intreal}). The goal of RPI is to reconstruct a discrete representation of the object $O_{ij}=O(i\Delta_x,j\Delta_x)$, using only the measured intensities $I_{ij} = |\tilde{E}(i \Delta_k, j \Delta_k)|^2$ and knowledge of the incident probe wavefield $P_{ij}$. In this notation, $\tilde{E}(k_x,k_y)$ refers to $\mathcal{F}\{E(x,y)\}$, the Fourier transform of $E(x,y)$ and $\Delta_x$, $\Delta_k$ are step sizes in real space and Fourier space, respectively.

To understand how a reconstruction of $O_{ij}$ is possible from a single diffraction pattern without the use of a finite support constraint, we consider the case where $O(x,y)$ is band-limited to a maximum frequency $k_o$ which is smaller than the highest frequency contained in the probe $k_p$. As the visualization of this process in Figure \ref{fig:exp:intfour} makes clear, when $k_o \ll k_p$, the final diffraction pattern will occupy a much larger region of Fourier space than the object itself does. As a result, the measured diffraction pattern can easily contain more measurements than there are independent degrees of freedom in a band-limited object, leading to a potentially well-defined inverse problem on the space of band-limited objects. This suggests the reconstruction strategy that we eventually pursue, where we impose an explicit band-limiting constraint on the object in lieu of a finite support constraint.

We additionally note that the map between the complex-valued object and the complex-valued wavefield at the detector plane is conveniently linear. As a result, the inverse problem can be expressed as a specific case of the generic phase retrieval problem \cite{Candes2013}. This connection lets us propose a limit on how high the band-limiting frequency $k_o$ of the original object can be as compared to the probe's maximum frequency $k_p$ before the inverse problem becomes fundamentally unstable. We note that for phase retrieval to be well posed, the number of intensity measurements (which scales with the area of the diffraction pattern) must typically be greater than four times the number of complex parameters in the object (which scales with the object's support in Fourier space) \cite{Balan2006,Bandeira2014}. In Supplementary Section 5, we show that in the context of our reconstruction algorithm, this requirement limits the applicability of RPI to objects with resolution ratios $R = \frac{k_o}{k_p}$ below approximately $0.94$. Later, we demonstrate experimentally that reconstructions on objects with at least $R\leq0.6$ are feasible in practice.

A practical advantage of this method is that the raw diffraction pattern itself provides convincing evidence that the underlying object meets the band-limiting requirement. This is because the diffraction pattern is derived from the convolution of the Fourier representations of the object and probe. The speckles in the diffraction pattern will therefore generically fill a region defined by the binary dilation of the probe's support in Fourier space with the object's. As a result, for a BLR probe like that shown in Figure \ref{fig:exp:blrfour}, $k_p+k_o$ can be estimated as the highest frequency at which the measured intensity remains above the noise floor. In other words, if the true object does not satisfy the band-limiting requirement, the diffraction pattern itself will almost always reveal this via diffracted intensity at frequencies which are simply too high to be accessed by a sufficiently band-limited sample. This fact further suggests that reconstructions of band-limited objects may benefit from randomized illumination even if no explicit band-limiting constraint is applied - as is the case in CMI. This idea, and it's limitations, are explored further in our numerical experiments.



Finally, a real-space view of the probe's structure further motivates the use of an explicit band-limiting constraint to perform reconstructions from RPI data. As seen in Figure \ref{fig:exp:blrreal}, a fine mesh of zeros interpenetrates the BLR illumination - a feature not shared with CMI modulators. Attempting to reconstruct an object at a finer resolution than the speckle size will therefore inevitably lead to pixels in the reconstructed object which are poorly constrained simply because they are weakly illuminated. If we explicitly apply a band-limiting constraint with $R<1$, every pixel in the object will be illuminated by at least one speckle. In this way, the band-limiting constraint alleviates this issue of weakly constrained modes at the same time as it dramatically improves the reliability of reconstructions. The limitations imposed on RPI by this issue, and their relationship to the beamstop diameter, are explored further in Supplementary Section 6.

\subsection{Reconstruction Algorithm}

\begin{figure}
\centerline{\includegraphics[width=\textwidth]{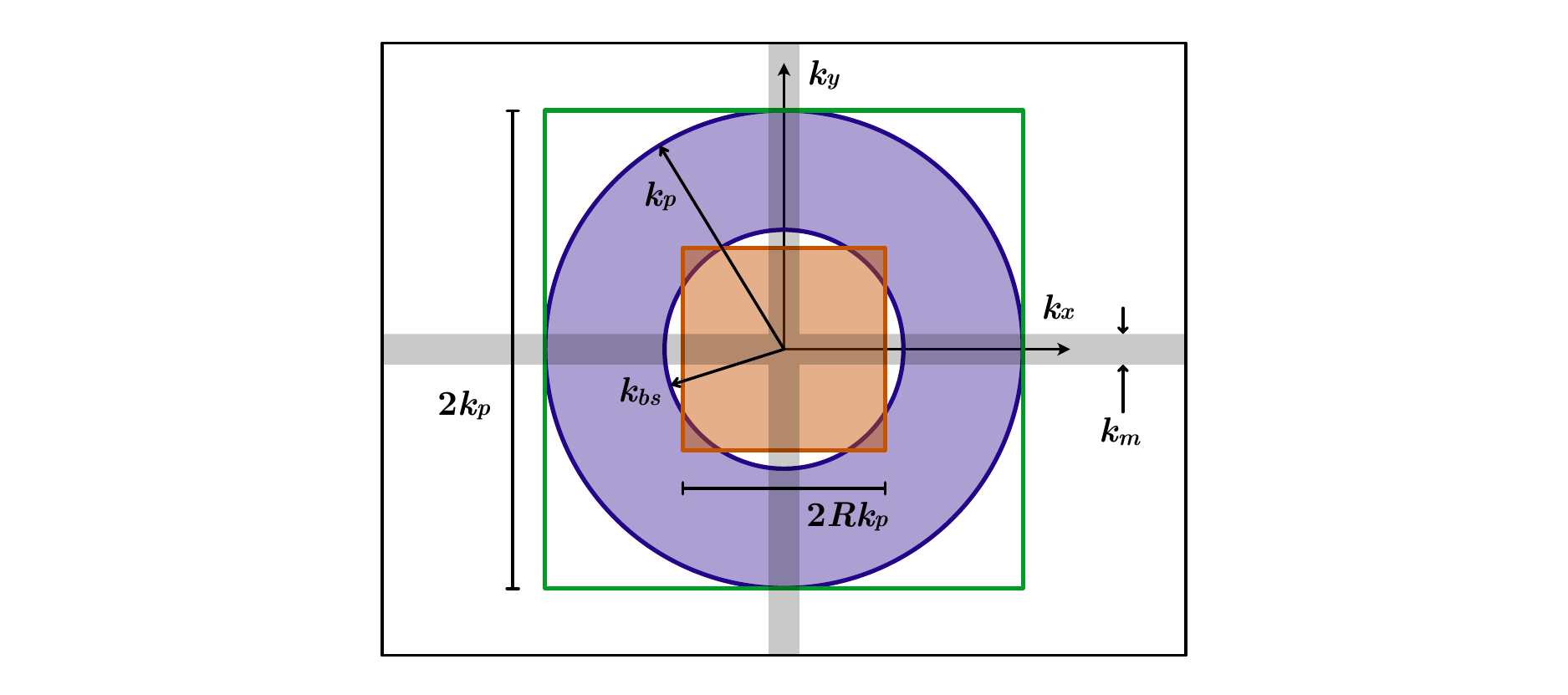}}
\caption{\textbf{Reciprocal Space Diagram}. The simulation geometry for our numerical experiments, as viewed in Fourier space. The purple region shows the extent of the probe function, bounded on the outside at a frequency $k_p$ and bounded on the inside by the extent of the central beamstop, $k_{bs}$. The orange region denotes the band-limited object, and the region bounded in green shows the extent of the noise introduced outside the band-limiting constraint. The grey shaded region is the extent of the simulated detector ``dead zone''.}\label{fig:num:diagram}
\end{figure}

Our reconstruction algorithm is is based on Automatic Differentiation Ptychography \cite{Ghosh2018,Kandel2019}, using the forward model defined in Equation \eqref{eq:forwardmodel}. This model incorporates the band-limiting constraint by taking a low-resolution representation of the unknown object, $O'_{kl}$, as input and using it in conjunction with a higher-resolution known probe $P_{ij}$ defined on the standard detector conjugate coordinate grid used in CDI and ptychography:

\begin{equation}\label{eq:forwardmodel}
\tilde{E}_{ij} = \mathcal{F}\{ P_{ij} \mathcal{F}^{-1}\{ \text{pad}(\mathcal{F}\{O'_{kl}\}) \}_{ij} \}.
\end{equation}

The first step in the model is to upsample the low-resolution object $O'_{kl}$ by padding it with zeros in Fourier space. We then directly multiply the upsampled object by $P_{ij}$, the high-resolution representation of the probe, and Fourier transform the resulting exit wave to propagate it to the detector plane. The size of the low-resolution object array $O'_{kl}$ is chosen ahead of time based on the measured radius of the freely propagated probe's diffraction pattern and a target resolution ratio $R_{rec}$.

To perform a reconstruction, we start with an initial guess of the object function and use Equation \eqref{eq:forwardmodel} to simulate the corresponding diffraction pattern. Next, we calculate the normalized mean squared error between the measured diffraction amplitudes and a simulated diffraction pattern (including a known detector background):

\begin{equation}\label{eq:lossfunc}
L = \frac{1}{\sum_{ij} I_{ij}} \sum_{ij} \left(\sqrt{|\tilde{E}_{ij}|^2 + B_{ij}} - \sqrt{I_{ij}}\right)^2.
\end{equation}

The quantity $L$ is referred to as the diffraction loss. We then use automatic differentiation to calculate the Wirtinger derivative of $L$ with respect to $O'_{kl}$ and feed those derivatives into an update for the object guess $O'_{kl}$ using the Adam algorithm \cite{Kingma2014}. This process is repeated iteratively until the object converges. Remarkably, we were able to use the same set of algorithmic parameters across all our numerical experiments as well as both experimental demonstrations, indicating the robustness of this reconstruction method to variations in experimental conditions.

Finally, we note that because the forward model simply defines a special case of the generic phase retrieval problem, other phase retrieval algorithms originally designed for geometries such as CDI and CMI can likely be modified to include a band-limiting constraint without affecting their validity. Potentially applicable families of algorithms include those designed to solve the generic phase retrieval problem, such as Wirtinger Flow \cite{candes2015} and SketchyGCM \cite{yurtsever2017}. In addition, algorithms such as hybrid input-output \cite{fienup1982} and difference map \cite{elser2003} that focus on finding solutions within the intersection of constraint sets are likely to be applicable, with the sets in this case generated by the band-limiting and detector intensity constraints. Although we pursued an Automatic Differentiation based approach here to provide the most flexibility while exploring this concept, it is likely the case that the use of tailored reconstruction algorithms such as these, designed specifically with the phase retrieval problem in mind, will lead to significant gains in computational efficiency.

\section{Results}

\subsection{Numerical Results}

\begin{figure}
\captionsetup[subfigure]{labelformat=nocaption}
\centerline{\includegraphics[width=\textwidth]{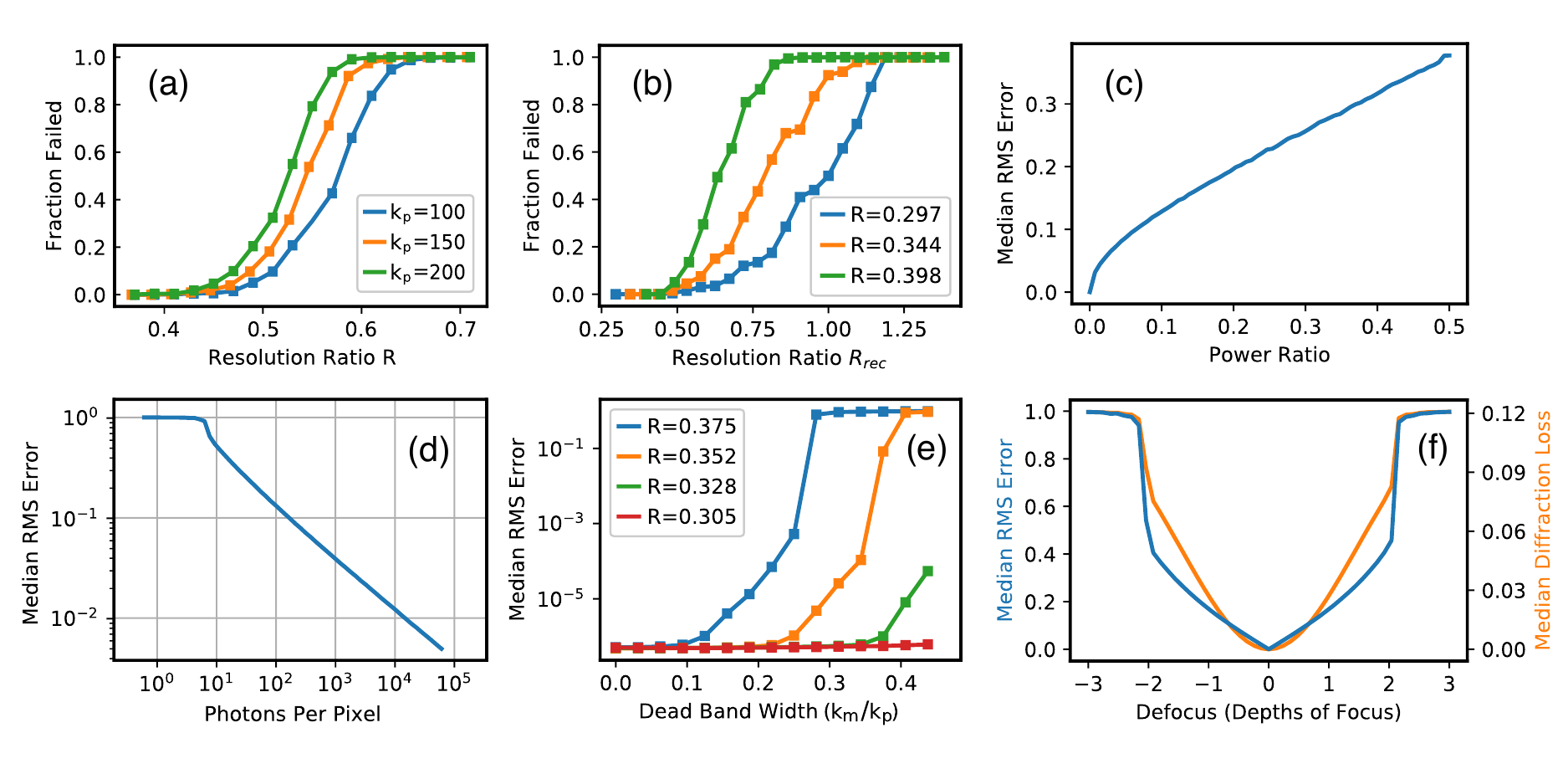}}
\begin{subfigure}{0\linewidth}
\caption{}\label{fig:num:resolution}
\end{subfigure}
\begin{subfigure}{0\linewidth}
\caption{}\label{fig:num:bandtight}
\end{subfigure}
\begin{subfigure}{0\linewidth}
\caption{}\label{fig:num:bandlimit}
\end{subfigure}
\begin{subfigure}{0\linewidth}
\caption{}\label{fig:num:poisson}
\end{subfigure}
\begin{subfigure}{0\linewidth}
\caption{}\label{fig:num:missingdata}
\end{subfigure}
\begin{subfigure}{0\linewidth}
\caption{}\label{fig:num:prop}
\end{subfigure}
\caption{\textbf{Numerical Results}. (a), The dependence of failure probability of a reconstruction on the true object's resolution ratio $R$ for various $k_p$ in pixels. (b), The dependence of failure probability on the resolution ratio $R_{rec}$ of the applied band-limiting constraint, for diffraction from underlying objects with various $R$. (c)-(e), The dependence of the median image RMS error $\epsilon$ (Eq. \eqref{eq:epsilon}) on (c), the ratio of power outside the band-limiting constraint to power inside; (d), the number of photons per illuminated pixel in the reconstructed object; and (e), the size of the cross-shaped dead region. (f), The dependence of the normalized diffraction RMS loss $L$ (Eq. \eqref{eq:lossfunc}) and the normalized image RMS error $\epsilon$ (Eq. \eqref{eq:epsilon}) on the error in the assumed propagation state of the probe.}\label{fig:numerical}
\end{figure}

Although the previous discussion demonstrates that RPI is valid for sufficiently small $k_o$, the precise region of parameter space where the inverse problem is well posed remains poorly defined. We therefore performed a collection of numerical experiments to clarify under what circumstances RPI reconstructions succeed, as well as to understand the impact of various noise sources on the quality and reliability of RPI reconstructions.

In the following numerical experiments, we used an implementation of our reconstruction algorithm developed in PyTorch \cite{NEURIPS2019_9015}, which is available from the authors upon request. All object guesses were initialized with Gaussian random real and imaginary parts, $\sigma=1$. The Adam reconstructions used $\alpha=0.4$, $\beta_1=0.9$, and $\beta_2=0.999$. For each reconstruction, $1000$ epochs of Adam were run, lowering $\alpha$ by a factor of $0.1$ whenever the diffraction loss $L$ (Eq. \eqref{eq:lossfunc})  failed to improve after $10$ iterations. Reconstructions were ended early if the loss fell below $1\times10^{-9}$ or $\alpha$ fell below $1\times10^{-4}$. The reported object error in each case is calculated from a region of the reconstructed object which falls entirely within the illuminated region. All reconstructions were run with a full detector, i.e. $k_m=0$ from Figure \ref{fig:num:diagram}, except those designed to uncover the impact of a finite missing region. In addition, unless otherwise noted, the beamstop diameter in Fourier space $k_{bs}$ is set to half the probe extent $k_p$, a typical choice which leads to good separation of diffraction orders in most common experimental geometries.

We first determined the maximum achievable resolution ratio $R=\frac{k_o}{k_p}$ in a noise-free experiment. To do this, we simulated diffraction from random objects with Gaussian-distributed real and imaginary parts, band-limited to different resolution ratios $R$. BLR illumination with $k_p = \{100, 150, 200\}$ pixels was used in the geometry described in Figure \ref{fig:num:diagram} and $1000$ RPI reconstructions were attempted for each diffraction pattern, using a target resolution ratio $R_{rec}=R$. Information on typical reconstruction results and the full set of parameters are found in Supplementary Section 1. We monitored the success of reconstructions using the normalized Root Mean Squared (RMS) error $\epsilon$ of the reconstructed images $R_{kl}$ as compared to the ground truth $O'_{kl}$:

\begin{align}
\epsilon &= \frac{1}{\sqrt{\sum \left|O'\right|^2_{kl}}} \sqrt{\sum{ \left|O'_{kl} - \gamma R_{kl}\right|^2}} \label{eq:epsilon},\\
\gamma &= \frac{\sum O'_{kl}R^\dagger_{kl}}{\sum \left|R_{kl}\right|^2},
\end{align}

where $R^\dagger_{kl}$ represents the complex conjugate of $R_{kl}$. We find that reconstructions typically converge to machine precision or entirely fail within the allotted number of iterations, as seen in Figure S1. We therefore classify a reconstruction as successful when $\epsilon < 0.1\%$ and unsuccessful otherwise. As shown in Figure \ref{fig:num:resolution}, when $R<0.4$, the reconstructions are virtually guaranteed to succeed, whereas when $R>0.6$, our algorithm becomes almost completely ineffectual. This should be compared to the theoretical limit of $R\approx0.94$.

We next considered how tight the band-limiting constraint needs to be, compared to the true band-limited frequency of the underlying object. This is a critical question because it addresses the extent to which reconstructions can succeed on data from band-limited objects, even if a loose band-limiting constraint (or no constraint) is applied. This is relevant because of the close connections between the data generated in RPI experiments and data arising from experiments, such as CMI, that also make use of high-frequency randomized phase structures. Viewed from the lens of RPI, CMI reconstructions of objects with band-limited spectra are analogous to RPI reconstructions where no band-limiting constraint has been applied. Despite that, the reconstruction may still benefit - particularly because the band-limited nature of the object is hinted at by the diffraction pattern itself. In this way, information about the object's support in Fourier space could potentially be available to the reconstruction algorithm implicitly, through the intensity constraint at the detector plane.

To separate the value of the explicitly imposed band-limiting constraint from the effect of the band-limiting frequency of the underlying object, we simulated diffraction from a set of ground truth objects which were band-limited to $R=\{0.3,0.35,0.4\}$. We next ran reconstructions with the reconstructed object band-limited to values of $R_{rec}$ ranging from the true $R$ of the original object up to $R_{rec}\approx1.25$. We then considered the success or failure of each reconstruction using the same criteria as above. The results in Figure \ref{fig:num:bandtight} show that, especially for objects with a low $R$, the band-limiting constraint can be significantly relaxed without affecting the reliability of the reconstructions. However, especially for objects with higher values of $R$, relaxing the constraint too far leads to a quickly increasing probability of failure.

This experiment supports the view that from a practical standpoint, the band-limiting constraint has a clear impact on the reliability of actual reconstructions. It also suggests that once the maximum stable resolution ratio is determined for a particular experimental system, it is safe to run all reconstructions on that system at the maximum resolution. Finally, it indicates that under favorable enough circumstances it is possible to perform RPI reconstructions using only the band-limiting constraint naturally imposed by the discretization of the detector conjugate coordinate space. This provides a strong link with CMI and provides an alternate perspective on the mechanism underlying CMI reconstructions.

We next considered the robustness of RPI to objects which contain some spectral weight beyond the band-limiting constraint imposed by the reconstruction algorithm. To do this, we divided reciprocal space into one zone within the band-limiting frequency $k_o = Rk_p$ and a second zone extending to $k_p$ (Fig \ref{fig:num:diagram}). We then simulated diffraction from randomly generated objects normalized such that the power in the outer zone was a specified fraction of that within the inner one. Reconstructions were performed with $R_{rec}=R$, and the ground truth was taken as the low-resolution representation of the object consisting of only frequency components within the band-limiting constraint. The results in Figure \ref{fig:num:bandlimit} demonstrate that even if the band-limiting constraint is not perfectly satisfied, for many applications the resulting error is tolerable provided the object's dominant length scale is above the pixel size of the low resolution object.

We then studied how efficiently RPI uses the information in shot noise limited experiments. We compared the RMS error from a series of reconstructions against the number of simulated photons per illuminated pixel in the low-resolution object. We find that information starts to be extracted at signal levels close to 10 photons per pixel, with the normalized RMS error $\epsilon$ falling below $1\%$ by the time signal levels reach $20,000$ photons per pixel (Fig \ref{fig:num:poisson}).

We subsequently examined the impact of missing data - for example, due to damaged regions or missing regions in a segmented detector. We modeled this situation by masking off a cross-shaped region of pixels of width $k_m$ (Fig \ref{fig:num:diagram}) representing a typical detector dead zone. We then performed reconstructions at several resolution ratios $R$ while varying $k_m$. The results confirm that the major effect of missing data is simply to lower the maximum stable resolution ratio $R$ (Fig \ref{fig:num:missingdata}). This behavior is similar to that of CMI as described in \cite{Zhang2016}, but contrasts with traditional CDI and FCDI, where missing data lead to poorly constrained regions of Fourier space and real space, respectively, as discussed in \cite{thibault2006}.

Finally, we investigated the effect of poor alignment between the object plane and the plane at which $P(x,y)$ is known. We find that, when no correction is applied to account for the probe's propagation, reconstructions succeed within roughly 2 depths of focus ($\text{DOF} = \frac{\lambda}{2\text{NA}^2}$) (Fig \ref{fig:num:prop}). Furthermore, within this region the observable diffraction loss $L$ is correlated with the true RMS error $\epsilon$ of the underlying image. This means that calibration of the probe defocus state is possible from each individual diffraction pattern, removing the need for precise shot-to-shot alignment.

\subsection{Optical Experiment}

\begin{figure}
\captionsetup[subfigure]{labelformat=nocaption}
\centerline{\includegraphics[width=\textwidth]{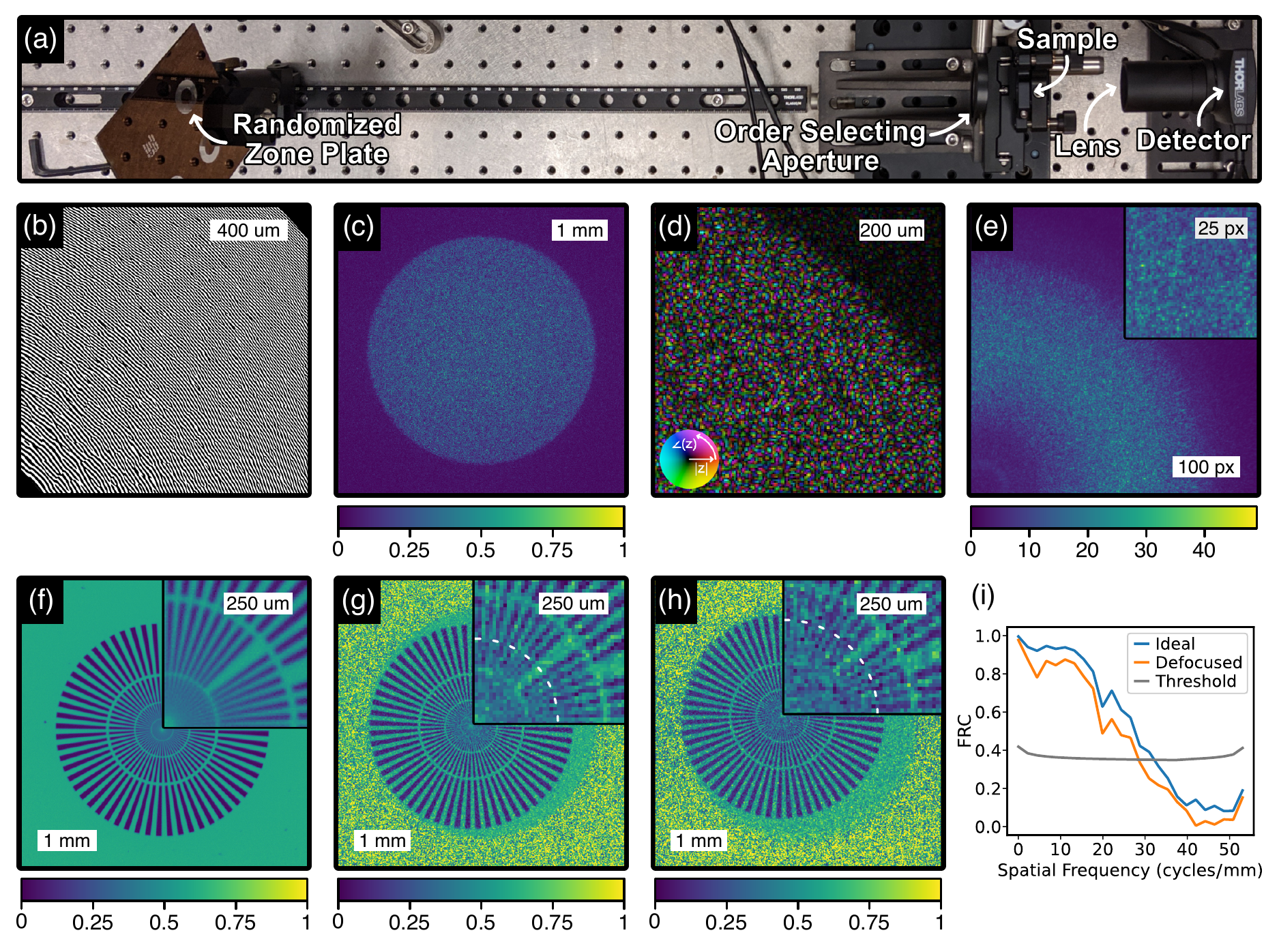}}
\begin{subfigure}{0\linewidth}
\caption{}\label{fig:opt:diagram}
\end{subfigure}
\begin{subfigure}{0\linewidth}
\caption{}\label{fig:opt:rzp}
\end{subfigure}
\begin{subfigure}{0\linewidth}
\caption{}\label{fig:opt:focalspot}
\end{subfigure}
\begin{subfigure}{0\linewidth}
\caption{}\label{fig:opt:focalzoom}
\end{subfigure}
\begin{subfigure}{0\linewidth}
\caption{}\label{fig:opt:pattern}
\end{subfigure}
\begin{subfigure}{0\linewidth}
\caption{}\label{fig:opt:ptycho}
\end{subfigure}
\begin{subfigure}{0\linewidth}
\caption{}\label{fig:opt:RPI1}
\end{subfigure}
\begin{subfigure}{0\linewidth}
\caption{}\label{fig:opt:RPI2}
\end{subfigure}
\begin{subfigure}{0\linewidth}
\caption{}\label{fig:opt:FRC}
\end{subfigure}
\caption{\textbf{Optical Demonstration}. (a), A top-down view of our optical test bench. (b), A section of the design pattern for the diffractive optic used in this experiment. (c), The amplitude of the incident probe, retrieved via ptychography. (d), A detailed view of the amplitude and phase of the recovered probe. (e), A segment of the square root of the measured diffraction pattern on the detector, in $\sqrt{\text{ADU}}$. Inset shows a further magnified region. (f), The amplitude of the test object, retrieved via ptychography. (g), The same test object's amplitude, retrieved from a single diffraction pattern via RPI. The dotted line shows the radius at which the pitch of the Siemens star matches the resolution calculated via FRC. (h), The amplitude of an RPI reconstruction from the same object after it was intentionally moved from the probe focus. (i), The Fourier ring correlation calculated between the ptychography and RPI reconstructions. All reconstructed amplitudes are reported in arbitrary units.}\label{fig:optical}
\end{figure}

To empirically validate RPI, we performed an experiment with a $532$ nm laser (Thorlabs CPS532) in a table-top setup. The experimental apparatus, shown in Figure \ref{fig:opt:diagram}, is a rough scale model of a typical scanning transmission X-ray microscope. The initial coherent wavefront was prepared using a beam expander with a $5$ $\mu$m spatial filter. This light then illuminates a $2$ cm diameter RZP (Fig \ref{fig:opt:rzp}) prepared via photolithography of a chrome-on-glass photomask. The RZP focused the light to a $4$ mm diameter spot at a focal distance of $33$ cm, using a design generated via the ``$\text{ZP}_0$'' method described in \cite{Marchesini2019} and discussed in Supplementary Section 2. Higher diffractive orders were removed through the combination of a $1$ cm beamstop integrated into the RZP and an iris placed approximately $2$ cm upstream from the sample acting as an OSA. The diameter of the iris was set to the smallest diameter at which no change in the freely-propagated first order diffraction pattern could be detected. A Siemens star test target was scanned through this focal spot to provide ptychography and RPI data.

The diffraction pattern was imaged on a monochrome camera (Thorlabs DCC1545M) with $5$ $\mu$m pixels in a $2f$ geometry using an achromatic doublet (Thorlabs AC254-150-A) with a $50$ mm focal length. The freely propagated probe fills roughly $\frac{2}{3}$ of the detector in this geometry. For the calibration ptychography scan, we collected a $20$x$20$ grid of diffraction patterns in $200$ $\mu$m steps (Figs \ref{fig:opt:focalspot}-\ref{fig:opt:ptycho}). Ptychography was performed on this data using Automatic Differentiation ptychography with a background correction, and the retrieved background was input into the RPI reconstructions.

We then performed an RPI reconstruction using a single diffraction pattern that was withheld from the calibration ptychography grid. The RPI reconstruction (Fig \ref{fig:opt:RPI1}) was performed using a $400$x$400$ pixel object, corresponding to $R_{rec}\approx0.6$ and a pixel size of $12.8$ $\mu$m. The parameters for the reconstruction algorithm were chosen to match those used in our numerical experiments, with the best result from a pool of $20$ random initializations reported in each case. The final single-frame full-pitch resolution of $31$ $\mu$m was calculated via a Fourier Ring Correlation (FRC) (Fig \ref{fig:opt:FRC}) between the RPI and ptychographic reconstructions at a threshold signal to noise ratio of $1$ \cite{VanHeel2005}, as discussed in Supplementary Section 3. This shows that studies of objects with resolution ratios of $R>0.5$ are possible in real experiments, with actual resolution nearly reaching the limit imposed by the band-limiting constraint.

Finally, to demonstrate the robustness of RPI we waited $4$ hours, power-cycled the illuminating laser, and defocused the probe by several millimeters before collecting an additional diffraction pattern. We then performed an ensemble of RPI reconstructions with computationally defocused probes and choose the result with minimum diffraction error. The retrieved defocus corresponded to a shift of the sample by $3.4$ mm along the propagation direction, roughly $12$ times the depth of focus of the zone plate. This reconstruction (Fig \ref{fig:opt:RPI2}) converged successfully but was found to have a slightly degraded full-pitch resolution of $35$ $\mu$m. 

These experiments show that RPI is straightforward to implement in a typical ptychography geometry. In addition, we show that the illumination function can be chosen to fill a high proportion of the natural detector conjugate space - in this case, almost $80\%$, without major issues. In other words, although the reconstructed object's resolution is lowered by the need to measure a larger region in Fourier space than that occupied by the band-limited object, there is a corresponding increase in the field of view because the requirement to sample the intensity distribution in Fourier space at the Nyquist rate (typically referred to as ``oversampling'' in the CDI literature) is relaxed. This leads to a final space-bandwidth product of the reconstructed object which is roughly equivalent to that of a standard CDI experiment using the same detector geometry.

\subsection{Soft X-ray Experiment}

\begin{figure}
\captionsetup[subfigure]{labelformat=nocaption}
\centerline{\includegraphics[width=\textwidth]{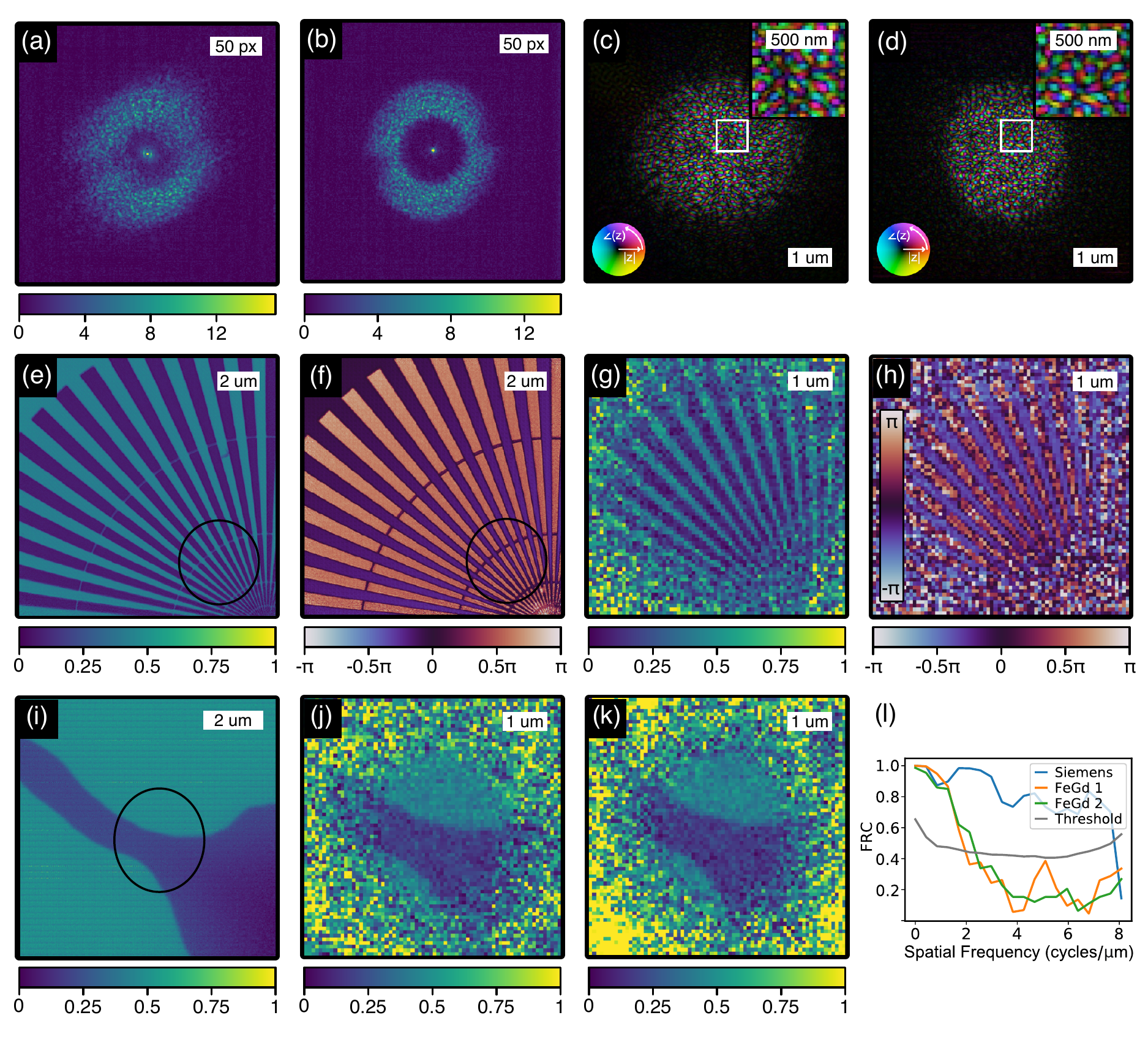}}
\begin{subfigure}{0\linewidth}
\caption{}\label{fig:sxr:sspattern}
\end{subfigure}
\begin{subfigure}{0\linewidth}
\caption{}\label{fig:sxr:fgpattern}
\end{subfigure}
\begin{subfigure}{0\linewidth}
\caption{}\label{fig:sxr:ssprobe}
\end{subfigure}
\begin{subfigure}{0\linewidth}
\caption{}\label{fig:sxr:fgprobe}
\end{subfigure}
\begin{subfigure}{0\linewidth}
\caption{}\label{fig:sxr:ssptyamp}
\end{subfigure}
\begin{subfigure}{0\linewidth}
\caption{}\label{fig:sxr:ssptyphase}
\end{subfigure}
\begin{subfigure}{0\linewidth}
\caption{}\label{fig:sxr:ssRPIamp}
\end{subfigure}
\begin{subfigure}{0\linewidth}
\caption{}\label{fig:sxr:ssRPIphase}
\end{subfigure}
\begin{subfigure}{0\linewidth}
\caption{}\label{fig:sxr:fgptycho}
\end{subfigure}
\begin{subfigure}{0\linewidth}
\caption{}\label{fig:sxr:fgRPI1}
\end{subfigure}
\begin{subfigure}{0\linewidth}
\caption{}\label{fig:sxr:fgRPI2}
\end{subfigure}
\begin{subfigure}{0\linewidth}
\caption{}\label{fig:sxr:FRC}
\end{subfigure}
\caption{\textbf{Soft X-ray Demonstration}. (a),(b), The square root of a diffraction pattern from the Siemens Star and FeGd samples, in $\sqrt{\text{ADU}}$. (c),(d), The probe as reconstructed from the Siemens Star and FeGd ptychography. Inset shows the detailed speckle structure. (e),(f), The amplitude and phase of the  ptychographic reconstruction of the Siemens Star. (g),(h), The amplitude and phase of an RPI reconstruction of a portion of the Siemens Star. (i) The amplitude of the ptychographic reconstruction of FeGd, showing three ferromagnetic domains. (j), The amplitude of an RPI reconstruction of a region of the Fe/Gd Sample. (k), The amplitude of an RPI reconstruction from the same diffraction pattern, using the probe reconstructed from the Siemens Star. (l), FRCs calculated for the various RPI reconstructions. All amplitudes are reported in arbitrary units.}\label{fig:softxray}
\end{figure}

Subsequently, we performed an experiment at the MAXYMUS beamline of BESSY II to demonstrate that RPI is applicable to X-ray microscopy. A $60$ $\mu$m diameter RZP was designed to focus $707$ eV light to a $2.6$ $\mu$m diameter focal spot over a focal length of $1.4$ mm, with a numerical aperture of $0.021$. Higher diffraction orders were filtered using a $30$ $\mu$m beamstop on the RZP and a $15$ $\mu$m OSA placed less than $385$ $\mu$m upstream from the sample. Scattered light was collected on a $264\times264$ CCD detector with $48$ $\mu$m pixels \cite{ordavo2011new} placed $17$ cm downstream from the sample.

We imaged a commercially available Siemens Star test sample (Carl Zeiss AG, Germany) made of 180 nm thick Au with a minimum feature size of $30$ nm, as well as a ferromagnetic Fe/Gd multilayer prepared by sputtering $50$ or $70$ alternating layers of  $3.6$ \AA\ thick Fe and Gd. All imaging was done at the Fe $L_3$-edge using circularly polarized illumination from the third undulator harmonic. The exit slits were set to 35$\times$35 $\mu$m, and a $200$ ms dwell time was used.

Typical raw diffraction patterns (Figs \ref{fig:sxr:sspattern}, \ref{fig:sxr:fgpattern}) show many of the issues common to coherent X-ray scattering experiments. A strong zeroth-order component from the zone plate is seen, which was masked off in our RPI reconstructions. The data also reveals damaged regions on the optic and astigmatism is visible in the reconstructed focal spots (Figs \ref{fig:sxr:ssprobe}, \ref{fig:sxr:fgprobe}). In addition, the ptychography reconstruction from the Fe/Gd multilayer (Fig. \ref{fig:sxr:fgptycho}) is affected by raster grid pathology.

Despite these shortcomings, single-frame RPI reconstructions were successful on both the Siemens star (Figs \ref{fig:sxr:ssRPIamp}, \ref{fig:sxr:ssRPIphase}) and Fe/Gd multilayer (Fig. \ref{fig:sxr:fgRPI1}). The reconstructions were obtained on a $70\times70$ pixel object with a pixel pitch of $83$ nm, corresponding to $R_{rec} \approx 0.5$. In both cases, the diffraction pattern used for RPI was withheld from the calibration ptychography reconstruction. The resolution was estimated via a FRC between the RPI and ptychography reconstructions (Fig. \ref{fig:sxr:FRC}). We find that the Siemens star, which contains strong high-frequency components, is reconstructed to the resolution of the pixel pitch at a signal-to-noise threshold of $1$, while the Fe/Gd sample is reconstructed at a full-pitch resolution of $510$ nm.

As a final demonstration, we performed a RPI reconstruction of the FeGd sample using a numerically propagated probe reconstructed from the Siemens star ptychography data (Fig. \ref{fig:sxr:fgRPI2}). We found that the probe's focal spot had shifted in the propagation direction by roughly $40$ $\mu$m between samples, roughly $20$ times the depth of focus of the zone plate. Ultimately, the retrieved full-pitch resolution of $420$ nm was actually negligibly improved.

These experiments again demonstrate the resiliency of RPI. Single-frame reconstructions succeeded despite serious aberrations in the optics and typical data quality issues. We also showed that the calibration ptychography and RPI reconstructions can be performed on different samples, with potential consequences for the ultimate resolution achieved. Therefore, only one X-ray pulse is required to interact with the sample of interest, potentially enabling ``diffract before destroy'' experiments.

\section{Discussion}

We have demonstrated a method for single-frame quantitative lensless amplitude and phase imaging that is reliable, straightforward to implement, robust to missing data, and applicable to extended samples. As a result, RPI complements and extends on many existing diffractive and traditional imaging methods. We therefore envision a variety of scenarios where RPI can become a valuable addition to the experimentalist's arsenal of techniques.

The most immediate use case we envision for RPI is as an alternative to off-axis holography, particularly in the soft X-ray regime. We note that while the resolution limits for both off-axis holography and RPI are comparable, in off-axis holography the signal rate diminishes as the size of the reference hole shrinks. Because RPI can make much more efficient use of the available flux, it is a preferable option for many experiments which would otherwise be performed using soft X-ray off-axis holography. As a bonus, because RPI doesn't require the deposition of a mask or the drilling of a reference hole, it is possible to study many systems with RPI which could not be prepared as samples appropriate for soft X-ray holography. 

RPI may also have value as an alternative to Transmission X-ray Microscopy (TXM), which it has two major advantages over. First, RPI uses every photon that hits the sample, while transmission X-ray microscopy uses an inefficient X-ray optic between the sample and detector. As light sources get brighter and brighter, sample damage thresholds increasingly limit the resolution of X-ray imaging experiments. In such a situation, moving an inefficient optic from downstream to upstream of the sample can be extremely valuable. The second advantage is that, whereas optical defects lead to aberrations in TXM images, all such aberrations are naturally corrected for in RPI during the probe calibration step. This could potentially lead to cheaper optics, larger optics, or even higher resolution optics than could be produced for a TXM experiment.

Similarly, RPI is an attractive alternative to CMI in many situations. Importantly, the experimental setup for RPI obviates the need for downstream optics in the sample's near-field region. This enables experiments in bulkier sample environments and removes a challenging alignment step. In addition, the RZP concentrates incident light onto the sample, allowing for much brighter illumination at the sample than can be achieved in a pinhole-based CMI geometry. The trade-off is the inevitable resolution limit that results from the truly band-limited nature of BLR illumination.

A second valuable application of RPI is as a complementary imaging method at transmission ptychography beamlines. In such a setup, users could switch between a high spatial resolution mode (ptychography) and a high time resolution mode (RPI) at will. This has obvious value - for example, allowing researchers to explore several regions of interest before commencing a dynamic imaging study, or making it possible to quickly study damage mechanisms in a material using RPI before commencing a static ptychography measurement. Although the large beam waist would prevent researchers from collecting fluorescence maps along with ptychography data, in many situations this trade-off would be valuable.

A third application of RPI is as a unique time-resolved quantitative imaging method for studying unpatterned thin films with soft X-rays in the reflection and Bragg geometries. Currently, no single-frame techniques exist for this kind of study. Because RPI is reference-free but much more reliable on highly textured samples, it is particularly well suited to this type of experiment provided that the sample itself is thin enough to ensure that the full angular frequency spectrum of the illumination can be diffracted.

Last, we propose a few avenues for improvement of RPI. First, our algorithm is tailored for flexibility. As a result, there are almost certainly gains in computational efficiency to be realized by porting other phase retrieval algorithms to this context. Second, it is possible that algorithmic improvements can push $R$ closer to the theoretical limit, improving the potential resolution of RPI experiments. Third, modern fabrication methods are capable of producing RZPs with higher numerical apertures than that used in our proof-of-concept experiment, further improving resolution. Finally, we note that the spatial resolution is limited by the numerical aperture of the optics, not the quality of the focal spot. This motivates the design of X-ray optics, such as those using higher diffraction orders, which are capable of generating BLR radiation at higher numerical apertures than would be possible if aberrations were a major concern.

\section{Conclusion}

We have demonstrated a reliable single-frame lensless X-ray imaging method that can be easily commissioned anywhere ptychography is done. As a result, it makes single-frame and time-resolved quantitative X-ray phase contrast imaging simpler, more reliable, and more accessible. RPI is likely to be applicable in fields where holography, TXM, 2D CDI, Fresnel CDI, ptychography, and X-ray photon correlation spectroscopy are used. Because of its flexibility and robustness, it will potentially enable a new generation of time-resolved studies on samples ranging from correlated materials and magnetic devices to soft matter, biological systems, and beyond.

\section*{Funding}
DOE Office of Basic Energy Sciences (DE-SC0019126);\\ National Science Foundation (NSF) (1751739);\\ STC Center for Integrated Quantum Materials (NSF DMR-1231319).

\section*{Acknowledgement}
We would like to acknowledge Bernd Ludescher of MPI-IS for preparing the FeGd sample and Prof. Gisela Sch{\"u}tz for providing infrastructure. We thank HZB for the allocation of synchrotron radiation beamtime at UE46-PGM2. We thank Stephan Hruszkewycz, Manuel Guizar-Sicairos, and Xiaojing Huang for providing valuable feedback on an early version of this manuscript. We would also like to thank Dmitry Karpov, Sujoy Roy, Colin Ophus, Steve Kevan, Flavio Capotondi, and Garth Williams for insightful discussions on the RPI concept. This work was supported by the DOE Office of Basic Energy Sciences, under Award Number DE-SC0019126, by the National Science Foundation under Grant No. 1751739, and by the STC Center for Integrated Quantum Materials (NSF Grant No. DMR-1231319).

\section*{Disclosures.}
The authors declare that there are no conflicts of interest related to this article.

See Supplement 1 for supporting content.

\bibliography{bibliography}

\end{document}


\maketitle

\section{Numerical Experiment Parameters}

\begin{figure*}
\centerline{\includegraphics[width=\textwidth]{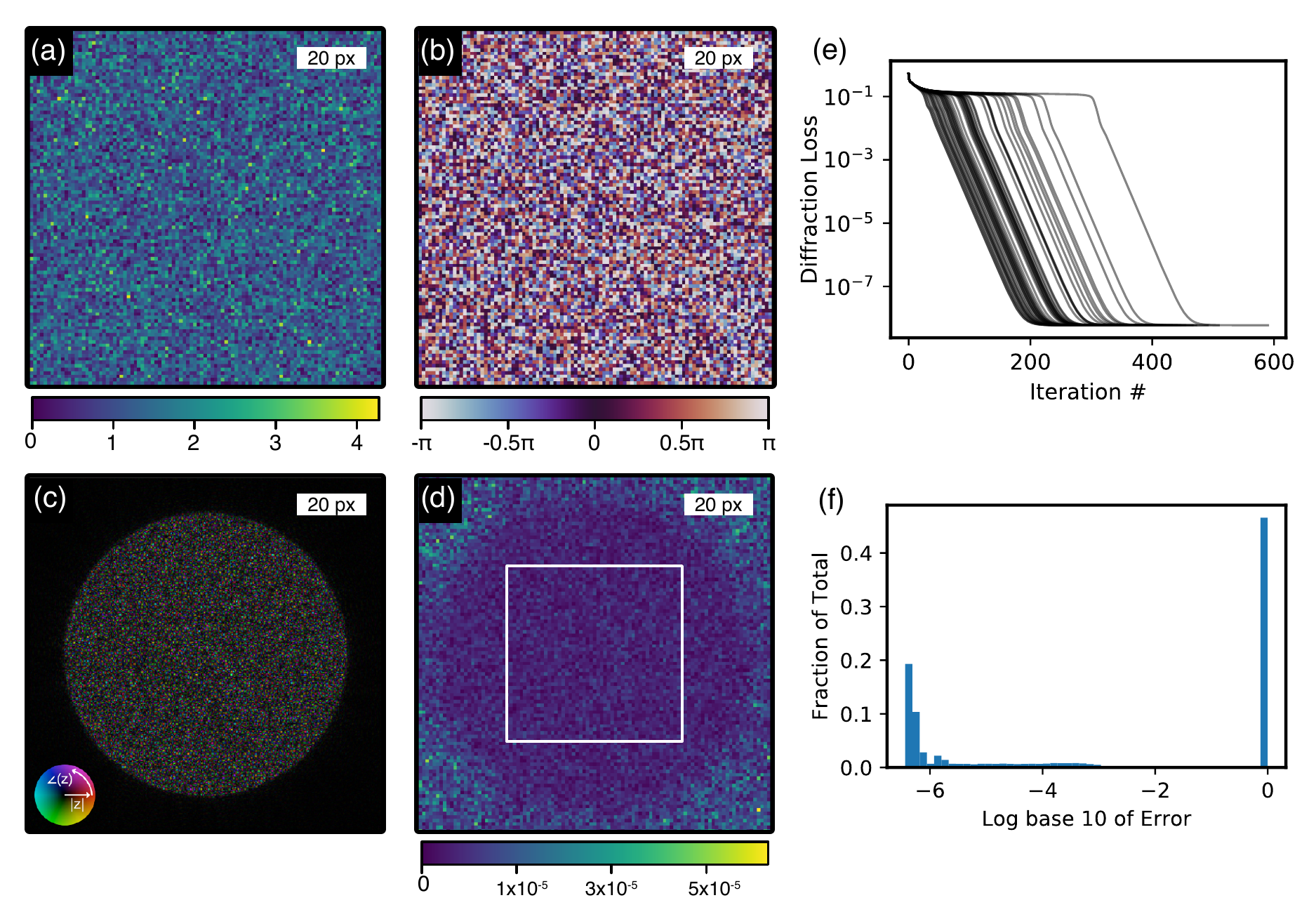}}
\caption{\textbf{Typical Reconstruction}. (a), (b) the amplitude and phase of a typical random object. (c), a typical randomized probe used in the simulations. (d), The magnitude of the difference between a completed noise-free RPI reconstruction and the ground truth, with the central region from which error is reported marked in white.  (e) The progression of diffraction loss over time for 50 independently initialized reconstructions on the same typical diffraction pattern. (f) The distribution of final RMS error for all reconstructions reported in Figure 3a.} \label{fig:typicalreconstruction}
\end{figure*}

We describe here the setup of our numerical experiments. Those interested in further inspecting the details of the experiments are encouraged to contact the authors for the original code.

In all cases, ideal BLR illumination was defined on an array with size larger than $2k_p\times2k_p$, where $k_p$ is the intended maximum probe frequency measured in pixels. To generate the illumination, a central circular region with a diameter $\frac{4}{5}$ of the array size is filled with uniform amplitude phase noise. This is then propagated into the far-field via a 2D Fourier transform. The region in Fourier space within $|k| = \frac{k_p}{2}$ is set to zero, as is the region outside $|k| = k_p$, leaving a ring in reciprocal space. This is then propagated back into the near field to form an ideal BLR probe for simulation.

Before simulating diffraction from the objects, the probes were upsampled by padding in Fourier space such that they are defined on an array of at least $(2k_p+2k_o)\times(2k_p+2k_o)$ pixels. This ensures that the multiplicative interaction between the probe and object doesn't lead to aliasing. We show the results of a typical ensemble of reconstructions from a randomly generated object in \ref{fig:typicalreconstruction}.

For the numerical experiments which discuss the impact of noise sources, all reconstructions were performed on a probe with maximum frequency $k_p=128$ and a band-limiting frequency of $R=0.4$. In each case, we performed reconstructions on between $50$ and $200$ randomly generated images per noise level. The error is determined from a square central region of the reconstructions, with a side length half that of the overall array, as shown in Figure S1d.

\section{Design of Randomized Zone Plates}

Our technique relies on illuminating a large region with highly speckled light containing high frequency components. This illumination function was achieved with diffractive optics that fill a clearly defined field of view with light whose intensity is as uniform as possible. The field of view is chosen to have a sharp drop-off simply so that the eventual reconstructions achieve comparable noise levels across the entire field of view. Other approaches which, for example, lead to a Gaussian envelope on the probe are equally applicable in principle, but lead to a less uniform noise profile in the reconstruction.

\begin{figure*}
\centerline{\includegraphics[width=\textwidth]{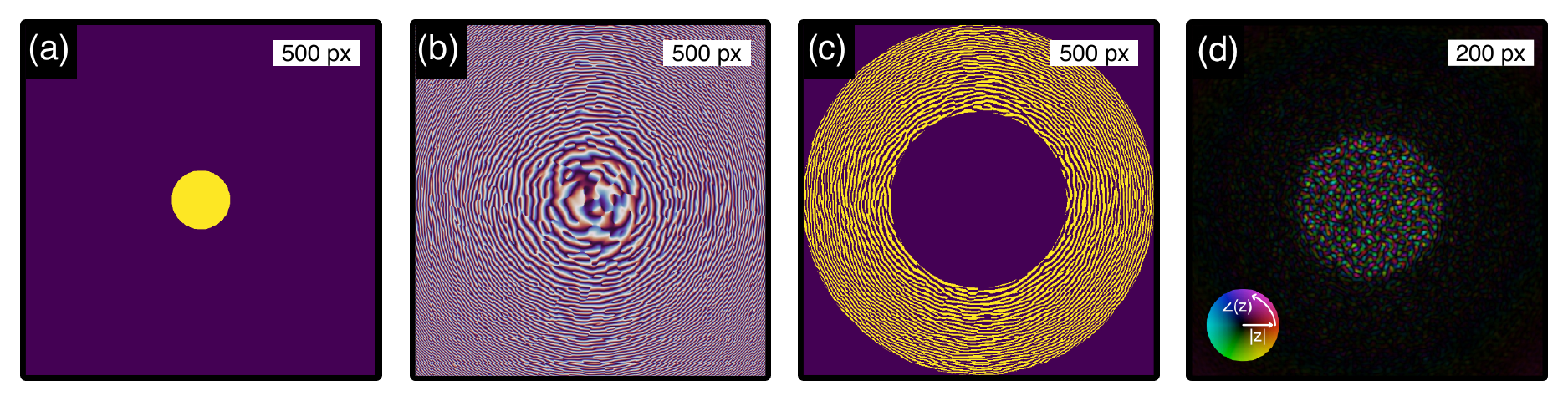}}
\caption{\textbf{Optics Design}. (a), The amplitude of the initial focal spot, filled with randomly varying phase. (b), The phase of this focal spot, propagated to the plane of the optic. (c), The binarized optic defined from this phase distribution. (d), The focal spot simulated from this binarized optic.} \label{fig:opticdesign}
\end{figure*}

To generate these diffractive optics, we start by simulating a discretized light field that consists of uniform amplitude pure phase noise within a circular aperture whose diameter matches the intended final field of view diameter. We then numerically propagate this light field to the plane of our diffractive optic, and record the phase of the light field at each pixel. We generate a binarized optic with a specified outer diameter and beamstop diameter from this phase field. The zones are defined by setting all pixels with phase below a certain threshold to $0$ and all other pixels to $1$. This process matches the simplest ``$\text{ZP}_0$'' approach described in detail in \cite{Marchesini2019}. This design process is further outlined visually in Supplementary figure \ref{fig:opticdesign}. All zone plates used in this paper used a beamstop with a diameter of half the outer zone plate diameter.

\section{Experimental Resolution Calculations}

All reported resolutions were calculated via the Fourier Ring Correlation (FRC)\cite{VanHeel2005} method, by comparison with ptychography results captured from the same sample region. Although this is potentially an underestimate of the true resolution if (as we saw in our reconstructions from FeGd samples) the ptychography itself exhibits pathologies, we believe that it is a conservative estimate which addresses the fundamental question of how RPI compares with ptychography.

In each case, we started by cutting out a region of the ptychography scan which overlapped with a region of the RPI reconstruction entirely within the RPI field of view. This ptychography cutout was then downsampled by extracting the central region in Fourier space corresponding to the band-limiting constraint in the RPI reconstruction. Any linear phase ramps in the RPI reconstructions (arising when the detector alignment shifts slightly between the calibration and RPI reconstructions) were manually removed and the cropped images were apodized with a Hann window. We then calculated a subpixel shift between the RPI and downsampled ptychography reconstructions and shifted the ptychography reconstruction to overlap with the RPI result. Finally, an FRC was calculated between the two images, excluding the outer 4 pixels to avoid artifacts from the circular shift of the ptychography reconstruction.

\section{X-ray Optics Fabrication}

The thin film stack was prepared by magnetron sputtering (Leica EM ACE 600, Germany) of a 100 nm thick Au film on a 100 nm thick SiN membrane with a window size of 500$\times$500 $\mu$m on a 200-$\mu$m thick Si frame (Silson ltd, UK). The ion beam lithography was done using a dualbeam focused ion beam instrument (Nova Nanolab 600, FEI, Netherlands) with a pattern generator attachment (ELPHY MultiBeam, Raith GmbH, Germany). The design bitmap pattern was converted to a dot map GDSII stream file and used as input. The binarization, using a 0.15:0.85 line-to-space ratio, resulted in a pattern that can be machined using a process that resembles a single-pixel-single-pass process discussed in \cite{Keskinbora2018}, and which resulted in the best pattern quality. The 60-$\mu$m wide computer-generated-hologram with 40-nm outermost width was milled using a 30-kV Ga+ ion beam and 50-pA current (19 nm nominal beam size) and 0.5-ms dwell time resulting in a dosage of 0.025 pC per dot.

\section{Calculation of Resolution Limit}

The theoretical limit on resolution is found by comparing the number of intensity measurements contained in the diffraction pattern $M$ to the number  of complex parameters in the object $N$. It is generally believed that, in order for the phase retrieval to be well posed, the number of intensity measurements must typically be greater than four times the number of complex parameters in the object \cite{Balan2006,Bandeira2014}. This  belief is conditioned on relatively weak assumptions regarding the diversity of information contained within the individual measurements. We note that $M$ is proportional to the area of the support of $\tilde{P}\ast\tilde{O}$ (the diffraction patttern). In contrast, $N$ scales with the support of $\tilde{O}$, with the same constant of proportionality. In our test geometry, we use a circular zone plate which fills a ring in reciprocal space $\frac{k_p}{2}<|k|<k_p$ with random phase noise. The band-limiting constraint applied to the object, however, is a square such that $|k_x|<k_o$,$|k_y|<k_o$.

For $k_o>\frac{k_p}{2}$, the support of $\tilde{P}\ast\tilde{O}$ will be a rounded square covering an area of $A_I = \pi k_p^2 + 8 k_p k_o + 4 k_o^2$. The support of $\tilde{O}$ will simply be $A_O = 4 k_o^2$. We can solve for the ratio, $R=\frac{k_o}{k_p}$, such that $A_I=4A_O$, finding $R\approx0.944$. In our paper, we chose to band-limit the object to a square region to allow for a simple computational approach and a well-defined interpretation of the resulting images. In a potentially more elegant reconstruction approach where the object is band-limited to a circular region in Fourier space, one finds the simpler result that the reconstructions are theoretically limited to $R=1$.

This result should be considered as a limit on the potential reconstruction resolution of RPI rather than an estimate of the likely achievable resolution. In order for the reconstruction problem to be well posed at $M=4N$, the measurement vectors must meet conditions which we do not guarantee. In addition, practical phase retrieval algorithms typically require more stringent constraints on the ratio of measurements to complex parameters to lead to convergence with high probability.

\section{Illumination Uniformity}

\begin{figure*}
\centerline{\includegraphics[width=\textwidth]{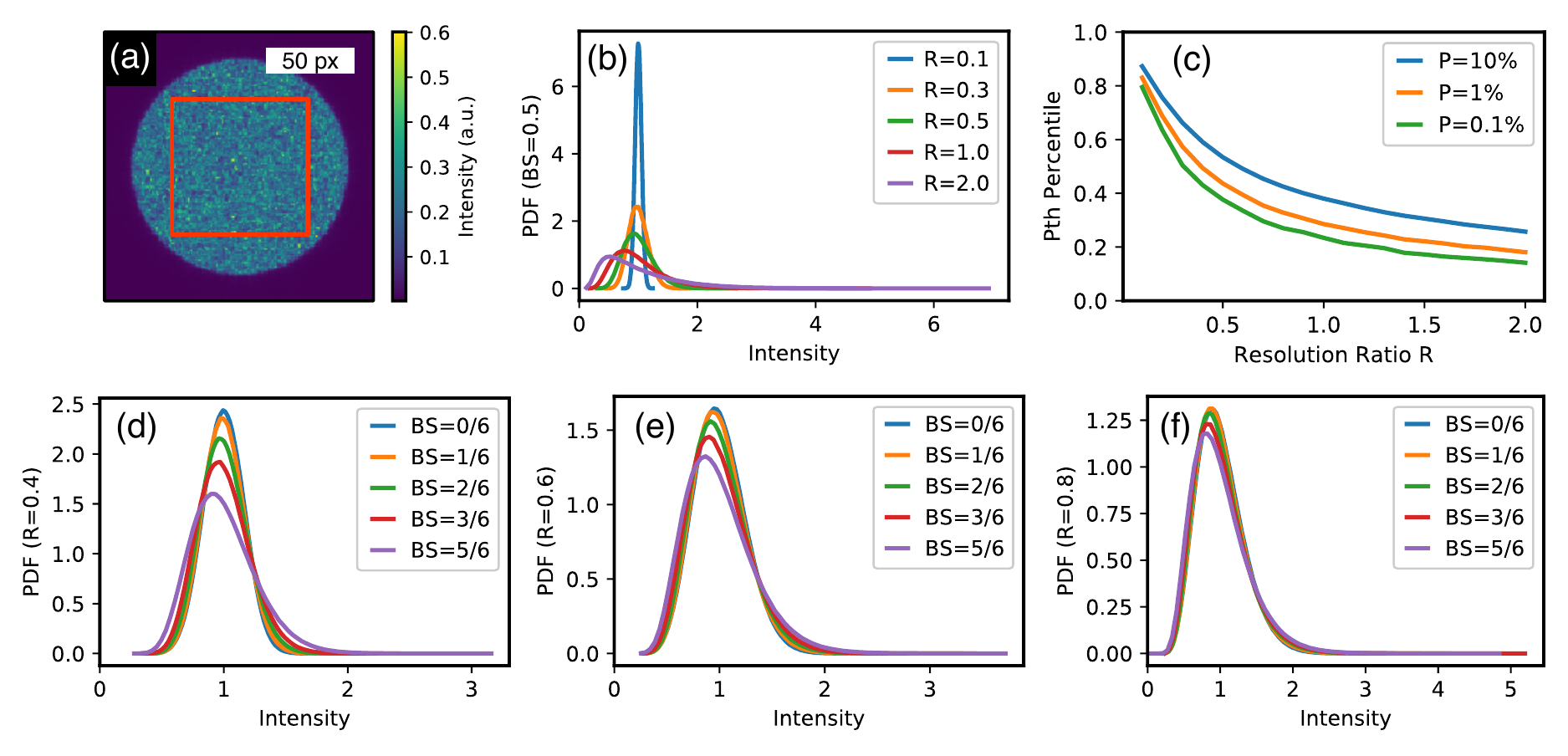}}
\caption{\textbf{Illumination Uniformity}. (a), The distribution of intensities over the focal spot for $R=0.6$, $BS=0.5$, with the region used to extract the PDFs outlined in orange. (b) The intensity PDF arising from BLR light with $BS=0.5$ at various resolution ratios. (c) The ratio of various percentiles to the median flux as a function of $R$ for $BS=0.5$. (d), (e), (f), The dependence of the illumination PDF on the beamstop diameter $BS$ at $R=0.4$,$R=0.6$, and $R=0.8$ respectively.} \label{fig:illuminationuniformity}
\end{figure*}

One issue which arises when using structured illumination is a variation in the intensity of the light which interacts with each pixel of the low-resolution object. In a shot-noise limited experiment, this naturally leads to higher uncertainty in the reconstruction of weakly illuminated pixels. Other noise sources which uniformly affect the diffraction pattern are also expected to lead to higher errors in the reconstruction of weakly illuminated pixels.

Conceptually, one can break the intensity variations into two distinct categories. Most important are the high-frequency variations that manifest themselves in the network of zeros interpenetrating the illumination. However, in addition to these nodes between the speckles, the speckles themselves have nonuniform intensities. This variation becomes important at high $R$ as the number of speckles per pixel decreases. Additionally, at large beamstop diameters correlations between neighboring speckles emerge. This reduces our ability to improve the uniformity by lowering $R$, because the various speckles within a low-resolution pixel remain correlated with one another.

To inspect the extent of this issue in RPI, we generated many BLR probes with $k_p=128$ at a variety of beamstop diameters using the framework outlined in Supplementary Section 1. The beamstop diameters are defined by $BS$, the ratio of the beamstop diameter to the optic diameter. Then, we ran the forward model on low-resolution objects at a collection of resolution ratios $R$. For each pixel in the object, we recorded the total diffracted intensity resulting from an object with that pixel set to 1 and all others set to 0. This measures how sensitive the final diffraction pattern is to the complex free parameter associated with that pixel. We then studied the distribution of resulting intensities arising from the pixels within the illuminated field of view.

The results are presented in Figure  \ref{fig:illuminationuniformity}. We find several important trends. The dominant trend is, as expected, an increase in the intensity variation with resolution ratio, $R$, at all beamstop diameters. In addition, illumination with a smaller diameter beamstop leads to smaller deviations at low $R$. This is a result of the beamstop-induced low-frequency intensity correlations. Nevertheless, we find that the distribution of intensities remains relatively small in the region $R<0.6$ where reconstructions are experimentally feasible. This remains true even when using a beamstop with the extraordinarily large diameter of $\frac{5}{6}$.

In conclusion, the illumination non-uniformity is likely to become an increasingly relevant issue if algorithmic improvements allow for numerically stable reconstructions at values of $R$ closer to $1$. However, for currently feasible reconstruction ratios and standard beamstop diameters, the variability in illumination is tolerable. To be specific, at $R=0.6$ and $BS=0.5$, $99.9\%$ of all pixels are illuminated by light which is at least $0.39$ times as intense as the flux through the median pixel.

\section{Application Recommendations}

First, we discuss recommendations for ptychography beamlines interested in adding a new zone plate to enable RPI. The most important parameter to consider is the numerical aperture of the randomized zone plate. In cases where the numerical aperture of the detector is large enough to enable reconstructions below a full pitch resolution of $20$ nm, the appropriate zone plate numerical aperture is likely to be ``as high as possible'', because it may not be practical to design a zone plate with a numerical aperture large enough to fill the detector. 

However, in a situation where it is possible to design such a zone plate with a numerical aperture matched to a detector, we recommend choosing a zone plate design which fills roughly $\frac{2}{3}$ of the detector at the lowest commonly used energy. This is because RPI reconstructions are likely to be reliable out to a resolution ratio of $R\approx0.5$. For a zone plate which fills $\frac{2}{3}$ of a detector, the highest reconstructed object frequencies at $R=0.5$ will be pushed exactly to the edge of the detector. One is tempted to increase the filling of the detector, however there is an advantage to leaving a portion of the detector unfilled. Because the high frequency components of an object are typically much weaker than the low-frequency ones, overfilling a detector with a high NA zone plate will swamp all the high frequency components with Poisson noise from the intense low frequency region. The portion of the diffraction pattern beyond the zone plate filling only includes high frequency components, and it is therefore desirable to capture it on the detector.

In all cases, we recommend designing the zone plate such that its focal spot fills roughly $80\%$ of the detector conjugate coordinate space at the highest commonly used energy. It may seem ideal to simply fill this space to get the most out of each pixel. An obvious downside is that, as the real space oversampling decreases, the finite pixel size on the detector leads to a reduced speckle contrast. In addition, including a small region around the focal spot allows some room for the probe to be defocused and therefore reduces the complexity of probe alignment. We finally note that, if a large energy range is used at a beamline, it may be worthwhile to design a collection of zone plates optimized for use at different energies.

We next discuss the issues that are relevant when designing a system from the ground up with RPI in mind. First, and most obviously, one must have set of motors capable of reliably scanning a test sample through the beam with accuracy high enough for ptychography to succeed. Second, there is an incentive to use a detector with as many pixels as possible. This is, unsurprisingly, because the space bandwidth product of the final RPI reconstruction will be related to the number of pixels in the detector. Fortunately, randomized zone plates lead to ptychography and RPI reconstructions which are robust to missing data, so there are no major issues with using segmented detectors to increase the pixel count.

If one is interested in the highest resolution imaging, it is prudent to design the highest numerical aperture zone plate which is reasonable, and ensure that the detector can subtend a numerical aperture which is at least $\frac{3}{2}$ times that. If one is interested in capturing lower resolution images with a larger field of view, it may be more appropriate to start by designing a detector geometry which leads to the desired field of view before designing a zone plate to match that detector's numerical aperture. Finally, designing with beamline stability in mind is especially critical, because a well characterized beamline would (for example) enable energy or polarization sweeps without the need for repeat calibration ptychography scans.

\bibliography{supp_bibliography}
